\documentclass{article}[final]

\usepackage[final]{neurips_2024}

\usepackage[utf8]{inputenc} 
\usepackage[T1]{fontenc}    
\usepackage{hyperref}       
\usepackage{url}            
\usepackage{booktabs}       
\usepackage{amsfonts}       
\usepackage{nicefrac}       
\usepackage{microtype}      
\usepackage{xcolor}         
\usepackage{graphicx}
\usepackage{mathtools}
\usepackage{enumitem}
\usepackage{wrapfig}

\usepackage[utf8]{inputenc} 
\usepackage[T1]{fontenc}    
\usepackage{hyperref}       
\usepackage{url}            
\usepackage{booktabs}       
\usepackage{amsfonts}       
\usepackage{nicefrac}       
\usepackage{microtype}      
\usepackage{xcolor}         

\newcommand{\mysection}[1]{
\vspace{8pt}\noindent
\textbf{#1.}
}

\title{Fundamental Risks in the Current Deployment of General-Purpose AI Models:\\What Have We (Not) Learnt From Cybersecurity?}

\author{%
  Mario Fritz \\
CISPA Helmholtz Center for Information Security\\ Coordinator of ELSA -- European Lighthouse on Secure and Safe AI\\
  \texttt{fritz@cispa.de} \\
}

\begin{document}

\maketitle


General Purpose AI - such as Large Language Models (LLMs) - have seen rapid deployment in a wide range of use cases. Most surprisingly, they have have made their way from plain language models, to chat-bots, all the way to an almost ``operating system''-like status that can control decisions and logic of an application. Tool-use, Microsoft co-pilot/office integration, and OpenAIs Altera are just a few examples of increased autonomy, data access, and execution capabilities.

\mysection{Cybersecurity Risk of Application Integrated General-Purpose AI Models}
Unfortunately, it turns out that the current technology is vulnerable to attacks like prompt and in-direct prompt injection. This means that a message sent to the AI by a user or even an attacker injecting a message into the AI, can alter the behavior and lead to malicious and harmful outcomes. The more powerful the AI's capabilities, the greater the potential harm.\\
Before the actually deployment in such scenarios, we have predicted such vulnerabilities and have hypothesized as well as demonstrated basically the whole range of known cybersecurity vulnerabilities that are induced by these issues  \citep{greshake23aisec} -- as illustrated figure.
In the meanwhile, this has been recognized as a core issue of the current technology that poses sever cybersecurity risks and has been reflected in threat taxonomies published e.g. by NIST and OWASP.
\vspace{-0.5cm}
\begin{figure}[h]
\centering
\includegraphics[width=0.7\linewidth]{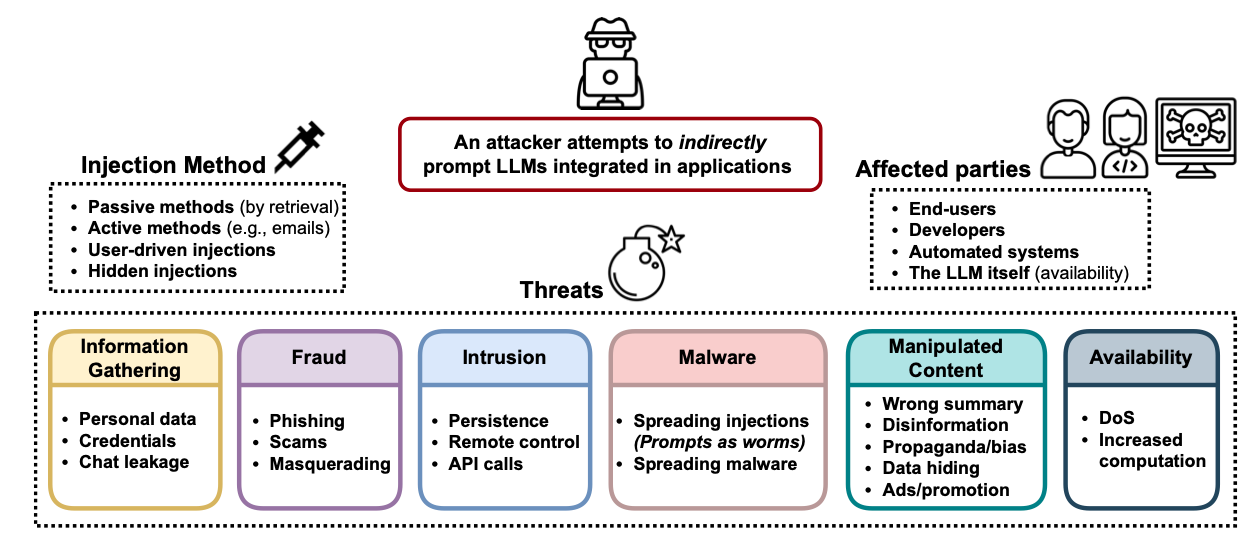}
\caption{Indirect Prompt Injection Threat Taxonomy \citep{greshake23aisec}.}
\end{figure}

\vspace{-0.5cm}
\mysection{Threats to Information Society by Shift in the Information Eco-System}\\
\vspace{-0.3cm}
\begin{wrapfigure}{r}{0.35\textwidth}
\centering
\includegraphics[width=0.9\linewidth]{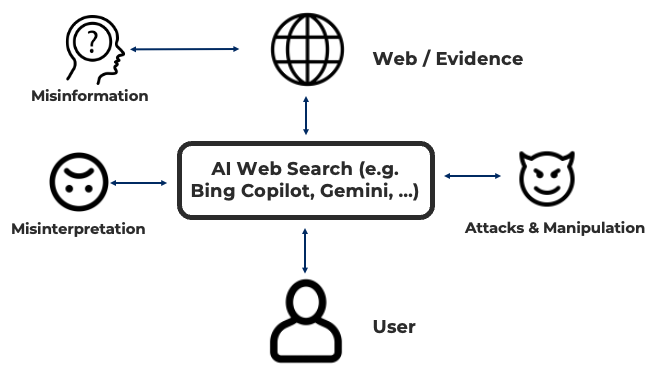}
\end{wrapfigure}
While our paper goes into great detail on these risks, it should be highlighted that the implications are very likely to be significant to the changing information ecosystem that in turn makes us even more vulnerable to disinformation. As illustrated in the figure to the right, we see a transition from consuming the ``raw'' information content on the web to increasingly relying AI assistants such as bing co-pilot or google Gemini to query and summarize information. This puts these information assistants in a central spot, as they are used to mediate the information and present it to us. While this setup might already by vulnerable to biases in the used model, the perspective of targeted attack on this new information architecture is even more concerning.

\mysection{Core Issue: Missing Data and Instruction Separation in General Purpose AI Models}
It turns out the core issues of such attacks like indirect prompt injection \citep{greshake23aisec}, is rooted in a fundamental cybersecurity requirement for IT system: Data-Instruction-Separation. As we argue and demonstrate \citep{zverev24setllm}, the current technology lacks such a separation and it is also not clear how to impose such constraints. Basically, a message sent to a general purpose AI / LLM can be interpreted as ``data'' or a new ``instruction'' that can drastically alter the behavior of the AI. 

We have formalized this issue and also proposed a {\it benchmark SEP} \citep{zverev24setllm} in order to quantify the capabilities of current LLMs to adhere to this essential Cybersecurity demands. Unfortunately, the current state of technology struggles with this issue - which makes it vulnerable when used in an application integrated context.

\mysection{ELSA Benchmarking Platform}
In the ELSA - European Lighthouse on Secure and Safe AI, we have worked quite extensively on benchmarking and auditing of AI models in particular w.r.t. robustness, privacy, and human agency/oversight. The ongoing benchmarks are available at \url{https://benchmarks.elsa-ai.eu}, which also features a related LLM-Capture-the-Flag challenge that was also hosted at the SATML conference \citep{llmctf24neurips}.

\mysection{Need for Mode Dynamic Test and Auto Red Teaming}
While most of the current benchmarks and auditing tools are ``static'', there is a clear need to make them more dynamic and adaptive. Relying on a secret AI validation set as the "root password to AI safety", seem highly dangerous. We would be running the risk of such tools being evaded and not representative of actual safety measures. In order to counter this,  we have taken inspiration from red teaming practices in cybersecurity to come up with with auto red teaming mechanisms that draw from prior experiences - similar to a ``red team play book''  \citep{chin2024incontextexperiencereplayfacilitates}. We also see opportunities to provide additional safety nets by detecting potential over-fitting - deliberately or indeliberately - to safety checks and benchmarks.

\mysection{Future Challenges of Multi-Agent Negotiations in General Purpose AI Models}
Another future direction, is the safety and reliability of multi-agent behavior. There is a lot of promise and interest to perform tasks like negotation and deliberation in a multi-agent setup, but we have shown for the first time, that these are equally vulnerable to adversarial behavior. Beyond this, we also show how to re-create tests, so that test contamination is reduced in order to arrive at more sustainable testing frameworks \citep{abdelnabi2024negotiation}. In addition, we also see an increasing need to evaluate the cybersecurity implications of AI code generation agents. We have contributed an initial benchmark that systematically finds vulnerabilities of state of the art code generation models \citep{hajipour24satml}. Lastly, in AI for Science, chemical and bio security of general purpose AI models are emerging challenges that are particularly difficult due the context dependent delineation between useful and harmful outputs.

\mysection{Auditing and Benchmarking vs. Foundational Approaches and Guarantees}
While benchmarking, testing, auditing play an important role, it has to be noted that this should not undermine the importance of foundational solutions and guarantees that rule out certain attacks and issues completely. Differential Privacy \citep{dp} and Robustness Certification \citep{li2023sok} are two excellent examples where research has progressed to a level that also for application relevant scenarios strong and rigorous statements about AI methods can be made. These statements hold and cannot be broken in the future. This is preferable - when ever possible - to empirical approaches.

\mysection{Vision for Secure and Safe AI in Europe -- A Strategic Research Agenda}
The ELSA - European Lighthouse on Secure and Safe AI has described a Strategic Research Agenda for Secure and Safe AI in Europe \citep{angelov_2024_10469287} that highlights challenges as well as opportunities. Key and strategic investments not only in the capabilities but also in the properties that will make AI comply with European values is a cornerstone of this joint vision.

\newpage
\mysection{Acknowledgments}
This work is supported in part by ELSA -- ``European Lighthouse on Secure and Safe AI'' funded by the European
Union under grant agreement No. 101070617.
Views and opinions expressed are those of the authors only and
do not necessarily reflect those of the European Union or European Commission. Neither the European Union
nor the European Commission can be held responsible for them.

\mysection{Note} This extended abstract was submitted to the European AI Office Workshop on  ``Evaluating General-Purpose AI Models with Systemic Risk''. It has received minor updates in order to reflect the actual presentation at the workshop.

\small
\bibliographystyle{abbrvnat}
\bibliography{references}

\end{document}